\documentclass[superscriptaddress, reprint, amsmath, amssymb, aps, prx, floatfix]{revtex4-2}

\usepackage[utf8]{inputenc}
\usepackage{graphicx}
\usepackage{dcolumn}
\usepackage{bm}
\usepackage[normalem]{ulem} 
\usepackage{mathrsfs}
\usepackage{physics}
\usepackage{xcolor}
\usepackage{amsmath, amssymb, mathtools, amsthm}
\usepackage[colorlinks=true]{hyperref}

\usepackage{comment} 
\usepackage{orcidlink}

\usepackage{xr}

\definecolor{ve}{RGB}{0,161,22}
\definecolor{dmag}{rgb}{0.6,0.0,0.6}
\definecolor{pink}{rgb}{1,0,0.9}

\begin{abstract}
Extracting momentum-resolved excitation spectra in strongly correlated quantum systems remains a major challenge, especially beyond one spatial dimension. We present an efficient tensor-network approach to compute dispersion relations via imaginary-time evolution within the infinite projected entangled-pair states (iPEPS) framework. 
Benchmarking on the transverse-field Ising model, the method successfully captures dispersion relations in both paramagnetic and ferromagnetic phases for two- and three-dimensional lattices, achieving strong agreement with series expansion methods, where these are applicable.
Crucially, this work presents the first demonstration of dispersion relation calculations for three-dimensional quantum lattice models---a long-standing computational challenge that opens entirely new research frontiers. The method demonstrates remarkable efficiency, requiring only modest computational resources while maintaining high accuracy across wide parameter ranges.
Its broad applicability makes it a powerful tool for quantum simulation, photonic material design, and quantum information platforms requiring precise momentum-resolved spectra.
\end{abstract}

\begin{document}

\author{Valeriia Bilokon\orcidlink{0009-0001-1891-0171}}
\email{vbilokon@tulane.edu}
\affiliation{Department of Physics and Engineering Physics, Tulane University, New Orleans, Louisiana 70118, United States}
\affiliation{Akhiezer Institute for Theoretical Physics, NSC KIPT, Akademichna 1, 61108 Kharkiv, Ukraine}

\author{Elvira Bilokon\orcidlink{0009-0007-8296-2906}\footnotemark[\value{footnote}]} 
\email{ebilokon@tulane.edu}
\affiliation{Department of Physics and Engineering Physics, Tulane University, New Orleans, Louisiana 70118, United States}
\affiliation{Akhiezer Institute for Theoretical Physics, NSC KIPT, Akademichna 1, 61108 Kharkiv, Ukraine}

\author{Illya Lukin\orcidlink{0000-0002-8133-2829}}
\email{illya.lukin11@gmail.com}
\affiliation{Akhiezer Institute for Theoretical Physics, NSC KIPT, Akademichna 1, 61108 Kharkiv, Ukraine}
\affiliation{Haiqu, Inc., 95 Third Street, San Francisco, California 94103, United States}

\author{Andrii Sotnikov\orcidlink{0000-0002-3632-4790}}
\email{a\_sotnikov@kipt.kharkov.ua}
\affiliation{Akhiezer Institute for Theoretical Physics, NSC KIPT, Akademichna 1, 61108 Kharkiv, Ukraine}
\affiliation{Karazin Kharkiv National University, Svobody Square 4, 61022 Kharkiv, Ukraine}

\author{Denys I. Bondar \orcidlink{0000-0002-3626-4804}}
\email{dbondar@tulane.edu}
\affiliation{Department of Physics and Engineering Physics, Tulane University, New Orleans, Louisiana 70118, United States}

\title{Dispersion Relations in Two- and Three-Dimensional Quantum Systems}

\maketitle
\footnotetext{These authors contributed equally to this work.}

\section{Introduction}
Probing the elementary excitations in quantum many-body systems remains one of the most fundamental challenges in condensed matter physics, with implications spanning from quantum phase transitions to quantum materials design. Dispersion relations serve as the fingerprint of emergent quantum phenomena, encoding the nature of quasiparticles that govern transport, phase transitions, and exotic quantum phases. From magnons in quantum magnets~\cite{Dyson1956, Holstein1940} to Cooper pairs in superconductors~\cite{Bardeen1957}, these momentum-resolved spectra reveal how microscopic interactions give rise to macroscopic quantum behavior.

The experimental accessibility of dispersion relations through angle-resolved photoemission spectroscopy~\cite{Damascelli2003, Sobota2021}, inelastic neutron scattering~\cite{Mourigal2013}, and terahertz spectroscopy~\cite{Mittleman2017, Hoffmann2011} has made them central to materials characterization and quantum technologies. Dispersion relations determine decoherence rates in quantum computing platforms~\cite{Maze2008}, guide quantum simulator design~\cite{Bloch2012, Gross2017}, and dictate the performance of spintronic devices~\cite{Zutic2004, Fert2008} and quantum sensors~\cite{Degen2017}. The ability to engineer dispersions through material design opens pathways to novel quantum phenomena, making accurate theoretical predictions essential for advancing the quantum frontier.

Despite their fundamental importance, calculating momentum-resolved excitation spectra in strongly correlated quantum systems presents formidable computational challenges. In particular, exact diagonalization~\cite{Avella2013, Dagotto1994} is limited to tiny clusters; quantum Monte Carlo~\cite{Avella2013, Dagotto1994, Troyer2005} suffers from sign problems in fermionic and frustrated spin systems; and series expansions~\cite{Oitmaa2006, Gelfand1990} have a limited convergence radius and diverge near critical points. While tensor network approaches such as matrix product states have delivered unprecedented accuracy in one dimension~\cite{Schollwock2011, White1992}, extending these successes to two and three dimensions~(2D and 3D, respectively) presents severe computational obstacles. The exponential growth of Hilbert space dimension with system size has thus created a computational bottleneck, leaving vast territories of quantum many-body physics—including most realistic models of quantum materials—beyond the reach of accurate dispersion calculations.

Infinite projected entangled pair states (iPEPS) have emerged as a promising approach to overcome these barriers through an efficient representation of ground states in two- and three-dimensional quantum lattice models~\cite{Jordan2008, Orus2014, Bruognolo2021, Corboz2021, Lukin2024single-layer3d}. Recent advances have demonstrated excitation spectra extraction using tangent space~\cite{Vanderstraeten2019, Tu2024} and corner transfer matrix methods~\cite{Ponsioen2020, Espinoza2024, Ponsioen2022}, with successful applications to paradigmatic models, such as the transverse-field Ising model (TFIM), Heisenberg model, and $J_1$-$J_2$ antiferromagnets. Nevertheless, progress has thus far been confined to two dimensions, and extending these capabilities to three-dimensional quantum systems remains an outstanding challenge.

In this work, we demonstrate the first systematic calculation of momentum-resolved dispersion relations for three-dimensional quantum lattice models using iPEPS, breaking through a long-standing computational barrier in quantum many-body theory. By extending the imaginary-time evolution approach to momentum-space observables, we extract complete dispersion relations across the Brillouin zone, benchmarking our method on TFIM in both two and three dimensions. Our results show excellent agreement with series expansion methods~\cite{Oitmaa2006, Gelfand1990} where available, while providing reliable spectra in parameter regimes where traditional approaches fail or have limited accuracy. Beyond benchmarking on the transverse-field Ising model, the approach is broadly applicable to a wide class of Hamiltonians, providing a powerful computational framework for quantum materials design, quantum simulator characterization, and fundamental studies of strongly correlated matter, where momentum-resolved excitation spectra encode the essential physics of emergent quantum phenomena.

\section{Extracting Dispersion Relation}

\subsection{Analytical Framework}
The calculation of dispersion relations in this study builds upon the spectral gap extraction method developed in Ref.~\cite{Lukin2024}. For a self-adjoint Hamiltonian $H$ with spectra $E_0 < E_1 < \cdots$ and a suitable-chosen self-adjoint observable $O$, the spectral gap $\Delta = E_1 - E_0$ can be extracted from the exponential decay of commutator expectation values during imaginary-time evolution:
\begin{align}\label{eq:spectral_gap}
    \ln\big|\bra{\phi(\tau)} [H,O] \ket{\phi(\tau)} \big| = -\tau\Delta + \mathcal{O}(1),
    \quad \tau \to \infty,
\end{align}
where $\ket{\phi(\tau)} = \mathcal{N}(\tau) e^{-\tau H}\ket{\phi(0)}$ is the normalized state propagated in imaginary-time from an initial state $\ket{\phi(0)}$. The observable $O$ should be chosen in such a way that it maps the ground state into the superselection sector with the quantum numbers of the excitation states of interest. 

To extract momentum-resolved excitation energies, we generalize this approach by introducing momentum-space observables. We perform a discrete Fourier transform of the local observable $O_j$ acting on site $j$:
\begin{equation}
    O_{\mathbf{k}} = \sum_{j} e^{i\mathbf{k} \cdot \mathbf{r}_j} O_j,
\end{equation}
where $\mathbf{k}$ is the momentum vector and $\mathbf{r}_j$ is the position vector of site $j$. Note that operators $O_{\mathbf{k}}$ map the ground state into the excitations with the fixed wavevector $\mathbf{k}$. Hence, the application of the spectral gap method to such observables should extract the spectral gap between the ground state and the lowest-energy excitations with the wavevector $\mathbf{k}$, while this is the definition of dispersion relations. 

Applying the spectral gap method with the momentum-space observable $O_{\mathbf{k}}$, Eq.~\eqref{eq:spectral_gap} becomes:
\begin{align}
    \mathcal{C}_{\mathbf{k}}(\tau) &= \ln\big|\bra{\phi(\tau)} [H,O_{\mathbf{k}}] \ket{\phi(\tau)} \big|, \label{eq:def_Ck}\\
    \mathcal{C}_{\mathbf{k}}(\tau) &= -\tau \Delta_{\mathbf{k}} + \mathcal{O}(1), \label{eq:disp_relation} 
\end{align}
where $\Delta_{\mathbf{k}} = E_1(\mathbf{k}) - E_0$ is the sought dispersion relation---the energy difference between the ground state and the lowest-energy excitation at momentum $\mathbf{k}$. In this way, the dispersion relation $\Delta_{\mathbf{k}}$ is extracted as the slope of the logarithmic decay in Eq.~\eqref{eq:disp_relation} for each momentum point $\mathbf{k}$.

\subsection{Tensor-network Methods}
The central step in the dispersion calculation is an imaginary-time evolution that projects the system into its ground state. For this purpose, we employ tensor-network techniques, specifically the iPEPS ansatz~\cite{Verstraete2008, Cirac2021, Bruognolo2021}. Starting from a randomly initialized iPEPS with bond dimension $D=1$, the state is optimized by successive imaginary-time steps within a periodic unit cell.  

We implement the evolution using two complementary approaches: matrix-product-operator (MPO) evolution and gate-based Trotterization (TG). In the MPO scheme (described in detail in Ref.~\cite{Lukin2024}), the exponential of the Hamiltonian is encoded in horizontal and vertical MPOs, which are sequentially applied to the iPEPS wave function. The bond dimension growth is controlled by superorthogonalization~\cite{Ran2012, Phien2015} and subsequent bond truncation with the help of superorthogonal canonical form.  

The gate-based approach, more common in practical iPEPS calculations, relies on a Suzuki–Trotter decomposition of the imaginary-time propagator. Unit cells of varying sizes can be employed to accommodate the structure of the Hamiltonian, with a minimal checkerboard unit cell containing two tensor types being the most common choice. The application of two-site gates increases the bond dimension $D$, which is truncated back to its target value using the simple-update method~\cite{Bruognolo2021, Jiang2008, Jahromi2019, Tindall2023}.  

With the tensor network evolution established, we proceed to extract the dispersion relation from the time-evolved states. After each time step $d\tau$, we evaluate the expectation value of the commutator $[H,O_{\mathbf{k}}]$ and plot $\mathcal{C}_{\mathbf{k}}(\tau)$~\eqref{eq:def_Ck} as a function of $\tau$. The dispersion relation is then extracted as a slope of the linear regime of the curve by a least-squares fit. Note that the expectation value of commutator is found with the approximate method based on simple-update environments \cite{Jahromi2019}. Although this approach is not exact, it still works well for gapped Hamiltonians. 

A crucial aspect of dispersion relation calculations within the iPEPS framework is the relationship between the unit cell size and momentum resolution. Due to the translational symmetry imposed by the periodic unit cell of the size $L_x \times L_y$~(in 2D) or $L_x \times L_y \times L_z$~(in 3D), the allowed momentum values are quantized according to
\begin{equation}
    k_i = 2\pi n_i / L_i,
\end{equation}
where $i \in\{x, y, z\}$, $n_i\in \{0, 1, 2, \dots, L_i-1\}$ and $L_i$ are the unit cell dimensions in lattice units. While finer momentum resolution requires larger unit cells, this comes at the cost of increased computational time for tensor network contractions. For the calculations presented in this work, we employ different unit cell sizes compatible with the desired momentum points. In particular, to access high-symmetry points we use a $2\times2$ unit cell in 2D (with the $\Sigma$ point evaluated on a $4\times4$ cell) and a $2\times2\times2$ unit cell in 3D.

\section{Results}

Let us consider the transverse field Ising model (TFIM) as a benchmark system. The latter is described by the Hamiltonian:
\begin{equation}
\mathcal{H} = -J \sum_{\langle i,j \rangle} \sigma_i^z \sigma_j^z - g \sum_i \sigma_i^x,
\end{equation}
where $\sigma_i^{x,z}$ are Pauli matrices at site $i$, $J>0$ is the Ising coupling strength, $g$ is the transverse field strength, and $\langle i,j \rangle$ denotes summation over nearest-neighbor pairs. This formulation applies to both square (2D) and cubic (3D) lattices. 

The model exhibits a quantum phase transition at a critical field strength $g_c/J$, separating a ferromagnetic phase ($g < g_c$) with spontaneous magnetization along the $z$-direction from a paramagnetic phase ($g > g_c$) where the ground state aligns with the transverse field. For the 2D square lattice, $g_c/J \approx 3.044$~\cite{Blote2002}, while for the 3D simple cubic lattice, $g_c/J \approx 5.29$~\cite{Garcia2013}. For our calculations, we adopt a phase-dependent parametrization: in the ferromagnetic regime, we fix the Ising coupling $J = 1$ and express the excitation gap in units of $J$. Conversely, in the paramagnetic regime, we fix $g = 1$ and express the excitation gap in units of $g$.

As discussed in Ref.~\cite{Lukin2024}, for the transverse-field Ising model, the operator $O = \sum_i \sigma^y_i$ is a suitable probe, since it connects the ground state to the first excited state for all values of $(g/J)$. The operator $O$ also satisfies the symmetry conditions required for spectral-gap analysis. In our case, to resolve the momentum-dependent excitations, we instead consider the Fourier transform of the corresponding operator
\begin{equation}
    O_\mathbf{k}= \sum_j e^{i\mathbf{k} \cdot \mathbf{r}_j} \sigma^y_j, 
\end{equation}
which yields the dispersion relation $\Delta_{\mathbf{k}}$.

\begin{figure}
    \centering
    \includegraphics[width=0.85\linewidth]{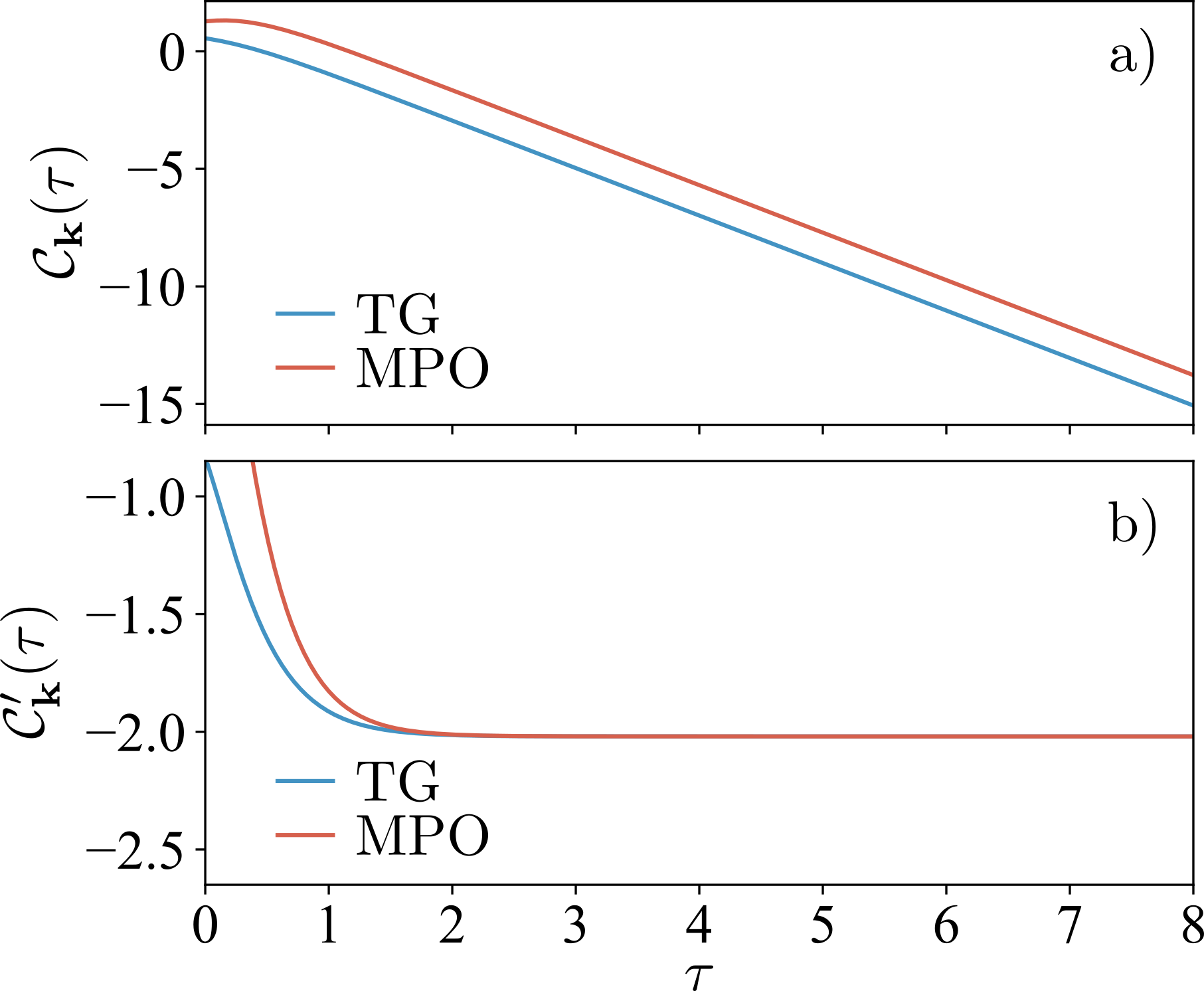}
    \caption{(a) Expectation value $\mathcal{C}_\mathbf{k}(\tau)$ and (b) its numerical derivative $\mathcal{C}_\mathbf{k}'(\tau)$ as functions of $\tau$ for the 2D TFIM in the paramagnetic regime with $J=0.1$ and $\mathbf{k}=(\pi, 0)$. The blue curve corresponds to data obtained by TG approach of the imaginary-time propagator, whereas the red curve represents the MPO scheme. For both methods, $D=5$~(see Appendix~\ref{app:convergence}) and $d\tau=0.01$.}
    \label{fig:sample}
\end{figure}

To demonstrate the accuracy and reliability of our iPEPS-based approach for extracting dispersion relations, we first examine the imaginary-time evolution of the commutator expectation value using two different computational methods: Trotter gates and matrix product operators.
Figure~\ref{fig:sample} illustrates the imaginary-time dynamics of the logarithmic expectation value $\mathcal{C}_\mathbf{k}(\tau)$ computed for 2D TFIM in paramagnetic regime with $J=0.1$. In Fig.~\ref{fig:sample}(a), both approaches converge to a linear asymptotic behavior at large imaginary times $\tau$, indicating exponential decay of the underlying commutator. This linear regime is clearly evidenced by the constant values observed in the derivative $\mathcal{C}_\mathbf{k}'(\tau)$ shown in Fig.~\ref{fig:sample}(b), where both curves reach a  plateau after some value of $\tau$. The slopes extracted from Fig.~\ref{fig:sample}(a) using linear regression provide direct access to the dispersion relation at $\mathbf{k}=(\pi,0)$, yielding $\Delta_\mathbf{k}=2.0170$ and $\Delta_\mathbf{k}=2.0197$ for the TG and MPO methods, respectively. The values closely agree with the value predicted by the perturbation series expansion~\cite{Oitmaa2006}, i.e., $\Delta_\mathbf{k}=2.0201$ (see also Appendix~\ref{app:expansions}). It validates the accuracy of both computational schemes and confirms the reliability of extracting spectral properties through imaginary-time evolution of commutator expectation values within the iPEPS framework. However, while the TG approach provides reliable results for moderate bond dimensions, it exhibits increasing instabilities and accuracy challenges as the bond dimension $D$ becomes large, making the MPO method more robust for high-precision calculations at large $D$.

\begin{figure}
    \centering
    \includegraphics[width=0.9\linewidth]{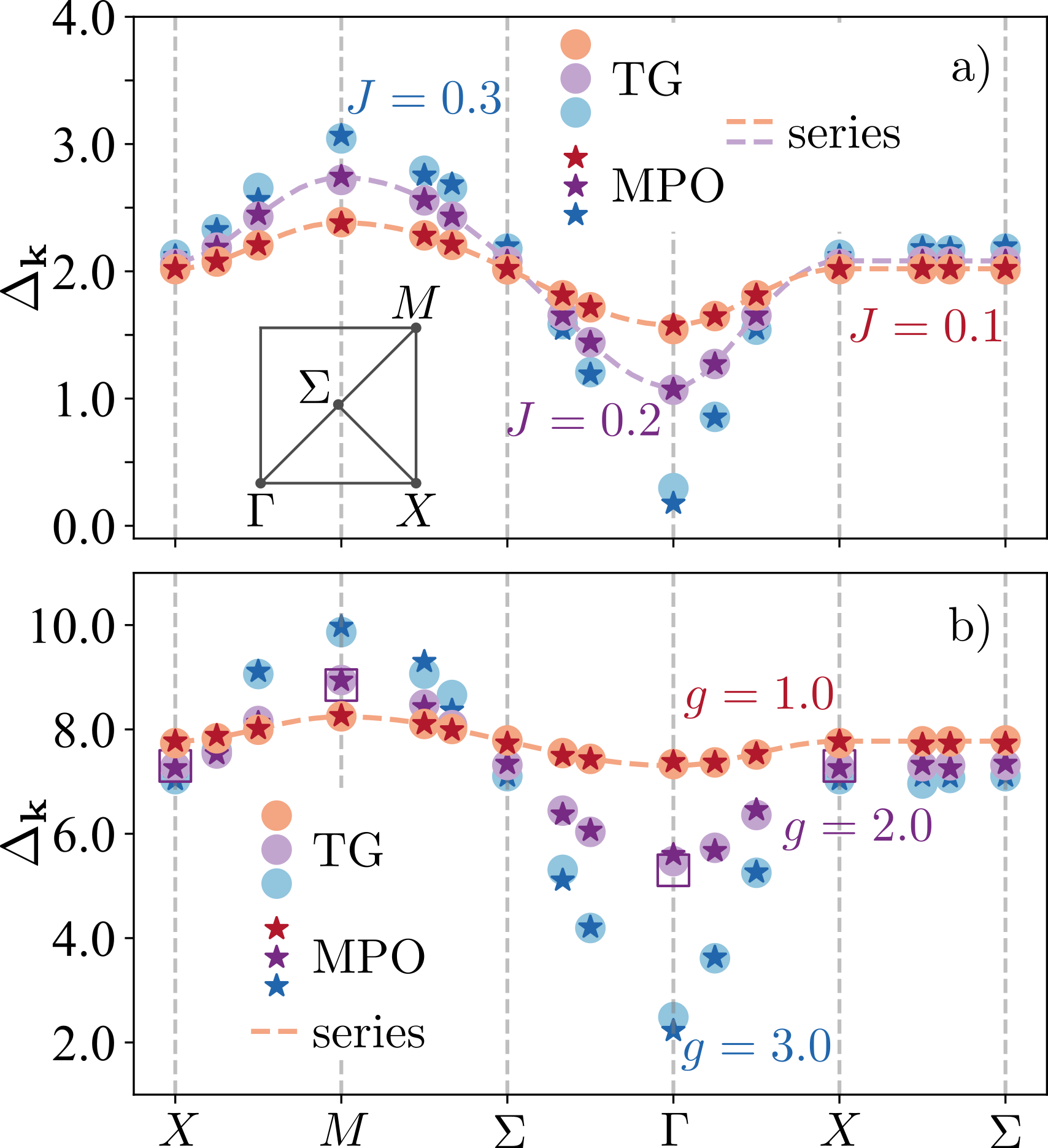}
    \caption{Dispersion relation for the 2D TFIM calculated using iPEPS with bond dimension $D=5$ and imaginary time step $d\tau=0.01$ in (a) paramagnetic and (b) ferromagnetic phases. Results are shown for different coupling strengths $J$ and field strengths $g$ as indicated. Circles represent Trotter gates (TG) calculations, stars represent matrix product operator (MPO) calculations, and dashed lines show dispersion relations obtained from series expansion~\cite{Oitmaa2006}. Squares indicate data points from Ref.~\cite{Espinoza2024}. The inset in panel (a) shows the Brillouin zone with high-symmetry points.}
    \label{fig:2d}
\end{figure}

Figure~\ref{fig:2d} presents the dispersion relation $\Delta_\mathbf{k}$ of the 2D TFIM along the path $X \!\to\! M \!\to\! \Sigma \!\to\! \Gamma \!\to\! X \!\to\! \Sigma$. In the paramagnetic regime [Fig.~\ref{fig:2d}(a)], shown for $J=0.1$, $J=0.2$, and $J=0.3$ with $g>g_c$, our iPEPS results closely follow the series expansion up to $\mathcal{O}(J^4)$~\cite{Oitmaa2006}. Figure~\ref{fig:2d}(b) illustrates the ferromagnetic regime for $J=1$ at $g=1$, $g=2$, and $g=3$. While the series expansion diverges at $g=2$, our results agree with the simulations from Ref.~\cite{Espinoza2024}.

At $\mathbf{k}=(0,0)$ the dispersion attains its minimum, corresponding to the gap of the system. We observe that calculations at the $\Gamma$ point become increasingly challenging as the field strength approaches the critical value $g_c$, where the convergence of our method slows significantly affecting the accuracy of the result. For parameters closer to criticality ($J=0.3$ in the paramagnetic phase and $g=3$ in the ferromagnetic phase), while the overall dispersion curves appear qualitatively correct, quantitative accuracy at the $\Gamma$ point decreases). Notably, the calculations at other high-symmetry points ($X$, $M$, and $\Sigma$) demonstrate excellent stability and maintain good agreement with the reference data. The accuracy of our simple-update-based approach across the Brillouin zone is further validated through comparison with a more advanced  tensor-network contraction scheme discussed in Appendix~\ref{app:CTMRG}. Together, these comparisons demonstrate that our approach reliably captures the spectrum across both phases.

\begin{figure}
    \centering
    \includegraphics[width=0.9\linewidth]{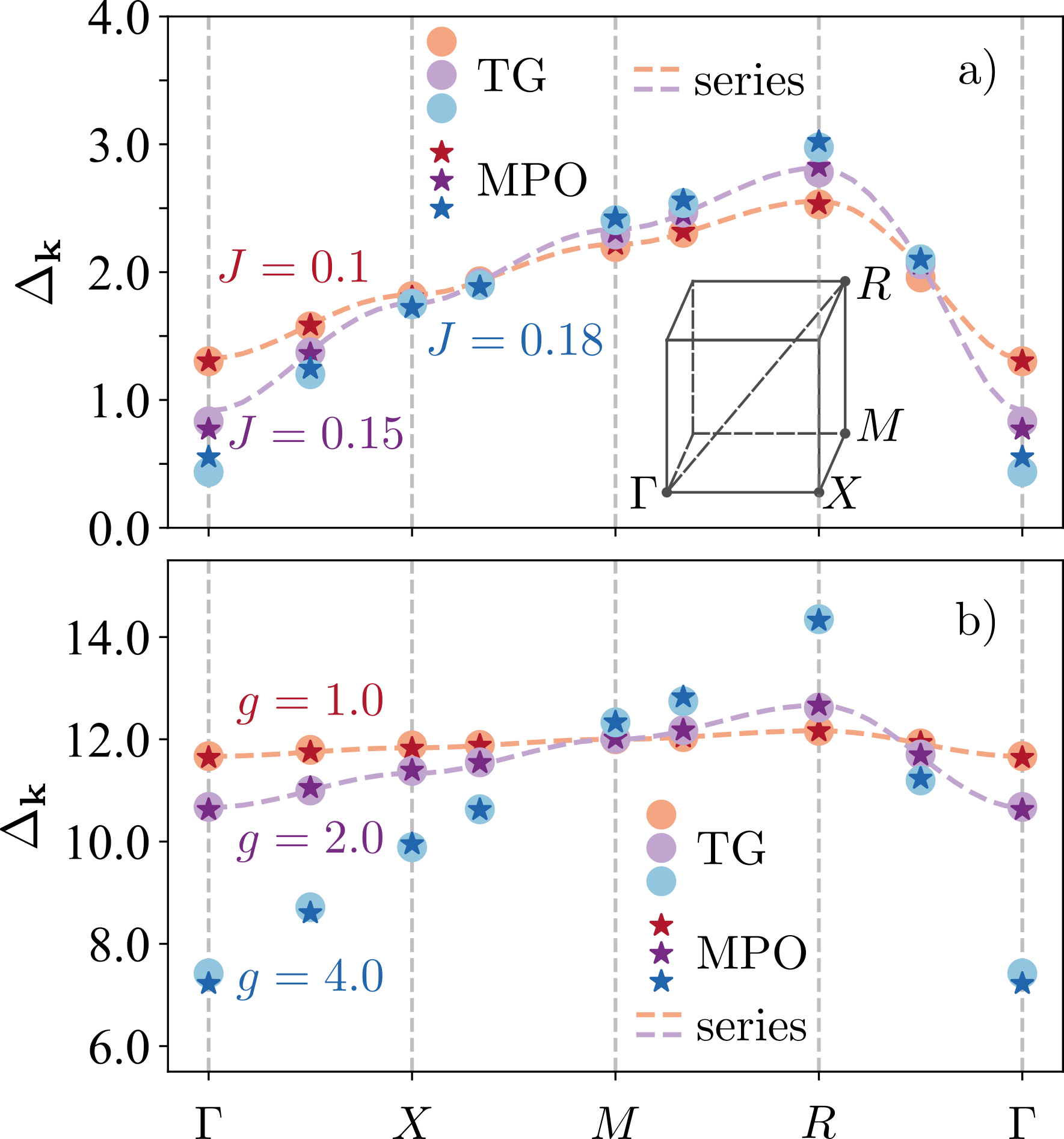}
    \caption{Dispersion relation for the 3D TFIM calculated using iPEPS with bond dimension $D=3$ and imaginary time step $d\tau=0.01$ in (a) paramagnetic and (b) ferromagnetic phases. Results are shown for different coupling strengths $J$ and field strengths $g$ as indicated. Circles represent Trotter gates (TG) calculations, stars represent matrix product operator (MPO) calculations, and dashed lines show dispersion relations obtained from series expansion~\cite{Oitmaa2006}. The inset in panel (a) shows the Brillouin zone with high-symmetry points.}
    \label{fig:3d}
\end{figure}

The extension to three dimensions presents significantly greater computational challenges due to the cubic tensor network structure. While PEPS/iPEPS approaches have yielded dispersion relations for 2D models in the thermodynamic limit, to the best of our knowledge, there is currently no published demonstration of dispersion relations for 3D quantum lattice models. Figure~\ref{fig:3d} displays the dispersion along the high-symmetry path $\Gamma \!\to\! X \!\to\! M \!\to\! R \!\to\! \Gamma$. In the paramagnetic regime [Fig.~\ref{fig:3d}(a)], our iPEPS results for $J=0.1$ and $J=0.15$ closely follow the series expansion~\cite{Oitmaa2006}, confirming the accuracy of our approach in three dimensions. 
For $J=0.18$, we present the first numerical calculation of momentum-resolved dispersion relations at this coupling strength.
In the ferromagnetic regime [Fig.~\ref{fig:3d}(b)], we show data for $g=1$, $g=2$, and $g=4$. For $g=1$ and $g=2$, our MPO and Trotter evolution results yield consistent agreement with series expansion~\cite{Oitmaa2006} across the Brillouin zone. At $g=4$, a regime inaccessible to series expansion techniques, we present the first numerical results for momentum-resolved dispersion relations. All simulations were carried out with bond dimension $D=3$ and time step $d\tau=0.01$. 
These results provide the first determination of momentum-resolved excitation spectra for the 3D TFIM and demonstrate that our method provides a reliable route to exploring higher-dimensional quantum systems.

\section{Conclusion}
We have introduced an efficient tensor-network framework for extracting momentum-resolved dispersion relations from imaginary-time evolution within the iPEPS ansatz. Demonstrated on the transverse-field Ising model as a benchmark system, our approach successfully captures dispersion relations across both paramagnetic and ferromagnetic phases in two and three dimensions. The results agree closely with the series expansions, validating the accuracy of the approach. Notably, to the best of our knowledge, this work presents the first demonstration of dispersion relation calculations for 3D quantum lattice models within the tensor-network computational approaches, thus opens new avenues for studying higher-dimensional quantum systems. The method remains accurate with modest bond dimensions and performs consistently with both Trotter and MPO evolution schemes across a wide range of coupling parameters, allowing all the simulations to be carried out on standard laptop-class hardware, without requiring access to high-performance computing resources.

The generality of the approach makes it broadly applicable to a wide class of lattice Hamiltonians, opening the door to systematic exploration of dynamical properties, transport phenomena in gapped strongly correlated quantum systems, as well as qualitative probes of spectral evolution near phase transitions. Our results have immediate implications for quantum simulation platforms, where understanding momentum-resolved excitation spectra is crucial for characterizing quantum phases and designing optimal protocols. The approach also provides new theoretical tools for photonic material engineering and quantum information applications, where dispersion properties directly impact transport phenomena and device performance. This work establishes a practical pathway for accessing the rich physics of momentum-dependent excitations in strongly correlated systems beyond the reach of conventional methods.


\acknowledgments

D.I.B. is supported by Army Research Office (ARO) (grant W911NF-23-1-0288; program manager Dr.~James Joseph). E.B. is supported by the National Science Foundation (NSF) grant No.~2403609. I.L. is supported by the U.S. National Academy of Sciences (NAS) and the Office of Naval Research (ONRG) through the Science and Technology Center in Ukraine (STCU) under STCU contract number [7120] in the framework of the International Multilateral Partnerships for Resilient Education and Science System in Ukraine (IMPRESS-U) and the IEEE program  ‘Magnetism for Ukraine 2024/2025’. A.S. acknowledges support by the National Research Foundation of Ukraine, project No.~2023.03/0073. The views and conclusions contained in this document are those of the authors and should not be interpreted as representing the official policies, either expressed or implied, of ARO, NSF, or the U.S. Government. The U.S. Government is authorized to reproduce and distribute reprints for Government purposes notwithstanding any copyright notation herein. 

We are grateful to Juan Diego Arias Espinoza and Philippe Corboz for sharing their numerical data for comparison purposes.

V.B., E.B., and I.L. contributed equally to this work.

\appendix

\section{Series Expansion Expressions for TFIM Dispersion Relations}\label{app:expansions}

This Appendix lists the explicit series-expansion expressions \cite{Oitmaa2006} used as benchmarks for the dispersion relations of the transverse-field Ising model in two and three dimensions. These high-order expressions, derived from linked-cluster methods, provide accurate analytical references for comparison with our iPEPS calculations.

In the paramagnetic phase the series are organized in powers of $J$, while in the ferromagnetic phase they are expanded in powers of $g$. We use the shorthand $c_{nx}=\cos(nk_x)$, with analogous definitions for $k_y$ and $k_z$. Below we list the leading terms.

\subsection{Paramagnetic Phase ($g=1$)}

The dispersion relation in the paramagnetic phase for 2D square~(SQ) lattice is given by:
\begin{align}
\Delta_\mathbf{k} &= 2 - 2J(c_{1x} + c_{1y}) + J^2\left[1 - 2c_{1x}c_{1y} - \frac{1}{2}(c_{2x} + c_{2y})\right] \nonumber \\
&\quad + \frac{1}{4}J^3\left[c_{1x} + c_{1y} - 6(c_{2x}c_{1y} + c_{1x}c_{2y}) - (c_{3x} + c_{3y})\right] \nonumber \\
&\quad + \frac{1}{32}J^4\bigl[70 - 16c_{1x}c_{1y} - 60c_{2x}c_{2y} - 24(c_{2x} + c_{2y}) \nonumber \\
&\quad - 40(c_{3x}c_{1y} + c_{1x}c_{3y}) - 5(c_{4x} + c_{4y})\bigr] + \cdots .
\end{align}

For the 3D case, the dispersion relation for simple cubic~(SC) lattice is:
\begin{align}
\Delta_\mathbf{k} &= 2 - 2J(c_{1x} + c_{1y} + c_{1z}) \nonumber \\
&\quad + \frac{1}{2}J^2\bigl[3 - 4(c_{1x}c_{1y} + c_{1x}c_{1z} + c_{1y}c_{1z}) \nonumber \\
&\quad -(c_{2x} + c_{2y} + c_{2z})\bigr] \nonumber \\
&\quad + \frac{1}{4}J^3\bigl[-24c_{1x}c_{1y}c_{1z} + (c_{1x} + c_{1y} + c_{1z}) \nonumber \\
&\quad - 6(c_{2x}c_{1y} + c_{1x}c_{2y} + c_{2x}c_{1z} + c_{1x}c_{2z} \nonumber \\
&\quad + c_{2y}c_{1z} + c_{1y}c_{2z}) \nonumber \\
&\quad -(c_{3x} + c_{3y} + c_{3z})\bigr] + \cdots .
\end{align}

\subsection{Ferromagnetic Phase ($J = 1$)}

The dispersion relation in the ferromagnetic phase in 2D for SQ lattice is:
\begin{align}
\Delta_\mathbf{k} &= 8 - \frac{1}{4}g^2(1 + c_{1x} + c_{1y}) + \frac{1}{768}g^4\bigl[19 + 12(c_{1x} + c_{1y})\bigr] \nonumber \\
&\quad - \frac{1}{884736}g^6\bigl[4745 + 4176c_{1x}c_{1y} + 4710(c_{1x} + c_{1y}) \nonumber \\
&\quad + 504(c_{2x} + c_{2y}) + 276(c_{2x}c_{1y} + c_{1x}c_{2y}) \nonumber \\
&\quad + 46(c_{3x} + c_{3y})\bigr] + \cdots. 
\end{align}

The ferromagnetic dispersion relation in 3D for SC lattice is given by:
\begin{align}
\Delta_\mathbf{k} &= 12 - \frac{1}{12}g^2(1 + c_{1x} + c_{1y} + c_{1z}) \nonumber \\
&\quad + \frac{1}{69120}g^4\bigl[151 + 60(c_{1x} + c_{1y} + c_{1z}) \nonumber \\
&\quad- 5(c_{2x} + c_{2y} + c_{2z}) \nonumber \\
&\quad - 20(c_{1x}c_{1y} + c_{1x}c_{1z} + c_{1y}c_{1z})\bigr] + \cdots .
\end{align}

\section{Convergence Analysis}\label{app:convergence}

\begin{figure}
    \centering
    \includegraphics[width=0.8\linewidth]{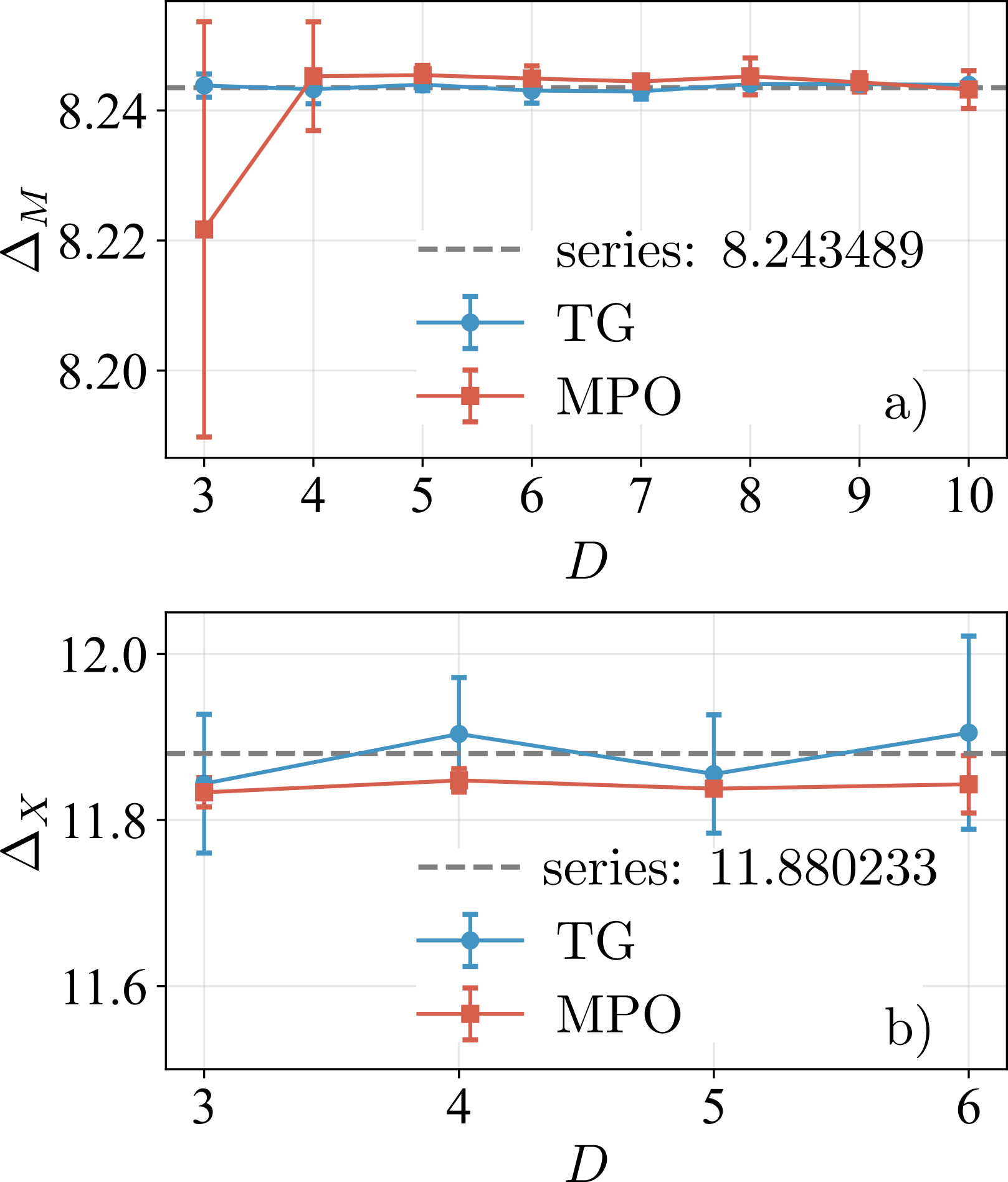}
    \caption{Convergence of the dispersion relation as a function of bond dimension $D$ for (a) 2D TFIM at the $M(\pi,\pi)$ point and (b) 3D TFIM at the $X(\pi,0,0)$ point. The system is in the ferromagnetic regime with $J=g=1$. Results from time evolution with Trotter gates (TG, blue circles) and matrix product operator (MPO, red squares) are compared against high-order series expansion values (dashed lines). Error bars represent the standard deviation across 5 independent trials for each bond dimension.}
    \label{fig:convergence}
\end{figure}

To validate the reliability of our tensor-network simulations, we benchmarked the single-particle excitation gap $\Delta_{\textbf{k}}$ at the high-symmetry momentum points: $M(\pi,\pi)$ in 2D and $X(\pi, 0, 0)$ in 3D. In Fig.~\ref{fig:convergence}, we show the convergence of $\Delta_{\textbf{k}}$ as a function of the bond dimension $D$ for two different implementations of the imaginary-time evolution: a Trotter-gate decomposition (TG) and a matrix-product-operator (MPO) representation. For each bond dimension $D$, we performed 5 independent trials to assess statistical variability. For references, we also include the results from high-order series expansions depicted in dashed lines.

For the 2D square lattice, we calculated the excitation gap $\Delta_M$ across bond dimensions $D\in[3, 10]$. Both approaches converge rapidly already at moderate values of $D\geq 4$. In Fig.~\ref{fig:convergence}(b), we analyze the convergence at the $X(\pi,0,0)$ point for bond dimensions $D\in[3, 6]$. Both TG and MPO methods demonstrate good agreement with the theoretical value $\Delta_X=11.880233$ at the bond dimensions studied.

In both dimensionalities, the methods systematically approach the benchmark value with increasing bond dimension. The rapid convergence and small residual fluctuations, together with agreement with the series expansion, provide strong evidence that our simulations capture the excitation spectrum with high accuracy. Importantly, the observed convergence indicates that the bond dimensions chosen in the main text are sufficient to reliably resolve the dispersion relation.

\section{Comparison with Full Contraction Scheme}\label{app:CTMRG}
\begin{figure}[h!]
    \centering
    \includegraphics[width=0.8\linewidth]{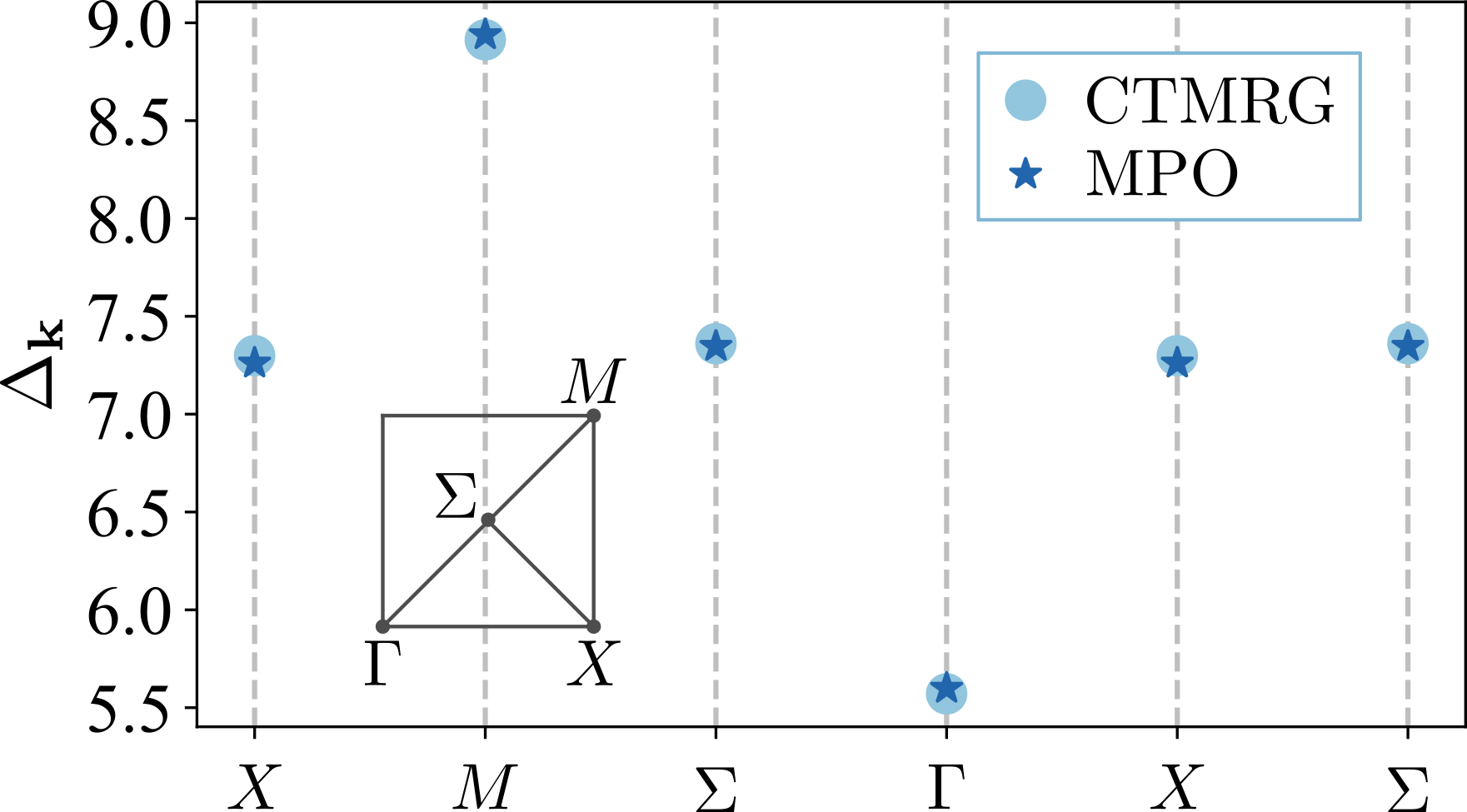}
    \caption{Dispersion relation for the 2D transverse-field Ising model in the ferromagnetic phase ($J = 1$, $g = 2$) calculated at high-symmetry points. Light blue circles represent results from CTMRG calculations (bond dimension $\chi=10$), and dark blue stars represent MPO evolution (bond dimension $D=5$) results. Imaginary time step is $d\tau=0.01$.}
    \label{fig:CTMRGvsMPO}
\end{figure}

To further assess the accuracy of the local contraction used in this study, we compare it against the more accurate corner transfer matrix renormalization group (CTMRG) method at $g=2$. Figure~\ref{fig:CTMRGvsMPO} compares the dispersion relations obtained using CTMRG (circles) and MPO with simple-update environments (stars) at all high-symmetry points. As shown, the excitation gaps extracted using both approaches agree very closely across the full high-symmetry path in the Brillouin zone. Differences are small and do not affect the qualitative or quantitative conclusions of the main text. This comparison confirms that for the gapped regimes studied in this work, simple-update environments provide sufficient accuracy for reliable dispersion relation extraction. However, we note that near critical points or in gapless phases, more sophisticated contraction schemes would become necessary. An important consideration is that such schemes would need to be applied at each time step of the imaginary-time evolution, not just for the final expectation value calculation, representing a substantial increase in computational cost. Note also that CTMRG was used only for observable calculation and not for the imaginary time evolution. The most reliable procedure of gap extraction near criticality requires also usage of CTMRG for time evolution itself. 

\bibliography{main}
\end{document}